\documentclass[10pt]{iopart}
\usepackage{bm}
\usepackage{amssymb}

\def\ba{{\bm a}}
\def\bb{{\bm b}}

\def\bn{{\bm n}}

\def\bx{{\bm x}}

\def\bz{{\bm z}}

\def\bN{{\bm N}}

\def\lb{\label}
\def\be{\begin{equation}}
\def\ee{\end{equation}}
\def\bea{\begin{eqnarray}}
\def\eea{\end{eqnarray}}

\begin{document}

\title[Time transfer functions]{General post-Minkowskian expansion of time transfer functions}

\author{Pierre Teyssandier\dag, and Christophe Le Poncin-Lafitte\dag\ddag}

\address{\dag\ D\'{e}partement Syst\`{e}mes de R\'{e}f\'{e}rence Temps et Espace, CNRS/UMR 8630\\
Observatoire de Paris, 61 avenue de l'Observatoire, F-75014 Paris, France}
\address{\ddag\ Lohrmann Observatory, Dresden, Technical University,\\
Mommsenstr. 13, D-01062 Dresden, Germany}

\begin{abstract}
Modeling most of the tests of general relativity requires to know the function relating light travel time to the coordinate time of reception and to the spatial coordinates of the emitter and the receiver. We call such a function the reception time transfer function. Of course, an emission time transfer function may as well be considered. We present here a recursive procedure enabling to expand each time transfer function into a perturbative series of ascending powers of the Newtonian gravitational constant $G$ (general post-Minkowskian expansion). Our method is self-sufficient, in the sense that neither the integration of null geodesic equations nor the determination of Synge's world function are necessary. To illustrate the method, the time transfer function of a three-parameter family of static, spherically symmetric metrics is derived within the post-linear approximation.
\end{abstract}


\maketitle

\section{Introduction}

\par In many tests of general relativity, it is crucial to explicitly know the function relating light travel time to the coordinate time of reception (or emission) and to the spatial coordinates of the emitter and the receiver. As the accuracy of measurements improves over the years, 
it will be soon indispensable to take into account the effects on the propagation of light predicted within the post-linear approximation 
of the relativistic theories of gravity. It will be the case, e.g., if experiments like the Laser Astrometric Test of 
Relativity (LATOR) \cite{turyshev} or the Astrodynamical Space Test of Relativity using Optical Devices (ASTROD) \cite{Ni} 
are carried out. 

In most of the studies devoted to determining light travel time, the 
relativistic effects are calculated by integrating the null geodesic equations (see, e.g., Refs. \cite{will} and \cite{blanchet}). General results 
have been obtained within the linear regime \cite{Kopeikin:1997,kopeikin2,kopeikin3,2003PhLA..308..101C}. However, the implementation of this method comes up against two problems. First, the integration of the null geodesic equations leads to heavy calculations in the post-post-Minkowskian approximation, even in the simple case of a static, spherically symmetric space-time \cite{richter1, richter2,john,brumberg}. Second, once the general solution is obtained, it is not easy to deduce from it the null geodesic path which joins an emitter and a receiver that have specified spatial positions. 

An alternative procedure enabling to overcome these two difficulties has been recently developed in \cite{2002PhRvD..66b4045L, 2004CQGra..21.4463L, universal}: the method consists in determining Synge's world function $\Omega(x_{A}, x_{B})$ for any pair of points-events $x_{A}$ and $x_{B}$ \cite{synge1}, and then deducing light travel time from the equation $\Omega(x_{A}, x_{B}) = 0$ which expresses that $x_{A}$ and $x_{B}$ are linked by a null geodesic path. This procedure works well, but presents 
an unpleasant drawback: once the general post-Minkowskian expansion of the world function is known, obtaining the corresponding expansion 
of light travel time still requires a lot of additional calculations (cf. Ref. \cite{universal}). It is the purpose 
of this paper to present a stand-alone method which totally avoids performing the calculation of the world function.  

\par The present work is organized as follows. In section 2 we summarize the notations and conventions used in the paper. In section 3 we recall the definition of the reception time transfer function and of the emission time transfer function, and we show that each of these functions obeys to a Hamilton-Jacobi-like partial differential equation. In section 4 we define the reception time delay function and the emission time delay function, and we show that each of these functions satisfies an integro-differential equation. Using this result, we derive in section 5 the general post-Minkowskian expansions of the time transfer functions. In section 6 we focus on the case of a stationary space-time. In section 7 we apply our method to the gravitational field of a static, spherically symmetric body treated within the second post-Minkowskian approximation. We present some concluding remarks in section 8.

\section{Notations}
\par In this work the signature of the Lorentzian metric $g$ is $(+,-,-,-)$. We suppose that space-time is 
covered by some global coordinate system $(x^{\mu})=(x^0,\bx )$. We put $x^0 = ct$, $c$ being 
the speed of light in a vacuum. Greek indices run from $0$ to $3$ and latin indices run from $1$ to $3$. Einstein 
convention on repeated indices is used here for expressions like $a^{i} b^{i}$ as well as for expressions 
like $A^{\mu} B_{\mu}$ or $a^{i}c_{i}$. Bold letters denote ordered triples. Given two triples $\ba = (a^1,a^2,a^3)=(a^i)$ 
and $\bb = (b^1,b^2,b^3)=(b^i)$, we use $\ba . \bb$ to denote $a^i b^i$. The quantity $\vert \ba \vert$ stands for the ordinary Euclidean 
norm of $\ba$: $\vert \ba \vert = (\delta_{ij}a^i a^{j})^{1/2}$. When it seems necessary for the sake of legibility, a quantity $f(x)$ is denoted by $(f)_{x}$ or $[f]_x$. The indices in parentheses characterize the order of 
perturbation. These indices are set up or down, depending on the convenience.

$G$ is the Newtonian gravitational constant.

\section{Time transfer functions}
\par We assume that space-time is globally regular with the topology of $\mathbb{R} \times \mathbb{R}^3$, that is without horizon. We suppose in addition that the coordinate system is chosen in such a way that the curves of equation $\bx = \bx_{A}$ are timelike for any given $\bx_A$, which means that $\partial /\partial x^{0}$ is a timelike vector field, i.e. $g_{00} >0$ everywhere. In agreement with these assumptions, we suppose that the past null cone at a given point $x_B = (ct_B, \bx_B)$ intersects the world line $\bx = \bx_{A}$ at one and only one point $x_A = (ct_A, \bx_A)$. The difference $t_B - t_A$ is the (coordinate) travel time of a light ray connecting the emission point $x_A$ and the reception point $x_B$. This quantity may be considered either as a function of the instant of reception $t_B$ and of  $\bx_A$, $\bx_B$, or as a function of the instant of emission $t_A$ and of $\bx_A$ and $\bx_B$. So it is possible to define two time transfer functions, ${\mathcal T}_r$ and ${\mathcal T}_e$ by putting\footnote{Note that the order of the arguments of ${\cal T}_{r}$ in equation (\ref{ttf}) is different from the one given in Refs. \cite{2004CQGra..21.4463L} and \cite{universal}.} 
\be \lb{ttf}
t_B - t_A = {\mathcal T}_r(\bx_A, t_B, \bx_B) = {\mathcal T}_e(t_A, \bx_A, \bx_B).
\ee

We call ${\cal T}_r$ the reception time transfer function and ${\cal T}_e$ the emission time transfer function. These functions are distinct except in a stationary space-time in which the coordinate system is chosen so that the metric does not depend on $x^0$.

Let us denote by $k^{\mu}$ the vector $dx^{\mu}/d\zeta$ tangent to the null geodesic path connecting $x_A$ and $x_B$, $\zeta$ being an arbitrary affine parameter of this geodesic. It was shown in \cite{2004CQGra..21.4463L} that the covariant components of $k^{\mu}$ satisfy the relations  
\be \lb{2d2}
\left(\frac{k_{i}}{k_{0}}\right)_{x_{A}} \, = \, c \, \frac{\partial {\cal T}_{r}(\bx_A,t_B,\bx_B)}{\partial x^{i}_{A}} 
\ee
and   
\be \lb{2d1}
\left(\frac{k_{i}}{k_{0}}\right)_{x_{B}} = 
-c \, \frac{\partial {\cal T}_{e}(t_A, \bx_A, \bx_B)}{\partial x^{i}_{B}}. 
\ee

The following theorem can be inferred from equations (\ref{2d2}) and (\ref{2d1})\footnote{Equations (\ref{eiko}) and (\ref{eiko2}) can also be derived from the general-relativistic version of Fermat's principle established in Ref. \cite{bel}.}.

\medskip
{\bf Theorem 1.} {\em The time transfer functions} ${\cal T}_{r}(\bx_A, t_B, \bx_B)$ {\em and} ${\cal T}_e (t_A, \bx_A, \bx_B)$ {\em satisfy the Hamilton-Jacobi-like equations}
\begin{eqnarray} 
\fl&& \qquad g^{00}(x^0_B - c{\cal T}_{r},\bx_A)+ 2c \, g^{0i}(x^0_B - c{\cal T}_{r},\bx_A)\frac{\partial {\cal T}_{r}}{\partial x^{i}_{A}} \nonumber \\
\fl&&\qquad \qquad \qquad \qquad \qquad \qquad + c^2\,g^{ij}(x^0_B-c{\cal T}_{r},\bx_A)\frac{\partial {\cal T}_{r}}{\partial x^{i}_{A}}\frac{\partial {\cal T}_{r}}{\partial x^{j}_{A}} = 0 \label{eiko}
\end{eqnarray}
{\em and}
\begin{eqnarray}
\fl&& \qquad g^{00}(x^0_A + c{\cal T}_{e},\bx_B)- 2c\,g^{0i}(x^0_A + c{\cal T}_{e},\bx_B)\frac{\partial {\cal T}_{e}}{\partial x^{i}_{B}} \nonumber \\
\fl&&\qquad \qquad \qquad \qquad \qquad \qquad+ c^2 \,g^{ij}(x^0_A+c{\cal T}_{e},\bx_B)\frac{\partial {\cal T}_{e}}{\partial x^{i}_{B}}\frac{\partial {\cal T}_{e}}{\partial x^{j}_{B}} = 0 , \label{eiko2}
\end{eqnarray}
{\em respectively}.

\medskip
{\bf Proof of theorem 1.} The covariant components of the vector tangent to $\Gamma_{AB}$ at $x_A$ satisfy the equation
\begin{equation} \label{equafonda}
\left(g^{\mu\nu}k_\mu k_\nu\right)_{x_A}= 0 .
\end{equation}

Dividing equation (\ref{equafonda}) side by side by $[(k_{0})_A]^2$, and then taking equation (\ref{2d2}) into account yield equation (\ref{eiko})\footnote{The covariant component $k_{0}$ of a null vector $k$ cannot vanish in the chosen coordinate system since the vector $\partial / \partial x^{0}$ is assumed to be timelike everywhere.}. The same  reasoning using $(g^{\mu\nu}k_\mu k_\nu)_{x_B}=0$ and equation (\ref{2d1}) leads to equation (\ref{eiko2}). 

In what follows, we give detailed proofs only for the reception time transfer function because similar procedures may be applied to the emission time transfer function.

\section{Integro-differential equations satisfied by the time delay functions}

Henceforth we suppose that the metric takes the form
\be \lb{pmm}
g_{\mu\nu} = \eta_{\mu\nu} + h_{\mu\nu}
\ee
throughout space-time, with $\eta_{\mu\nu} =$ diag$(1, -1, -1, -1)$. Then the contravariant components of the metric may be decomposed as
\be \lb{pmc}
g^{\mu\nu} = \eta^{\mu\nu} + k^{\mu\nu},
\ee
the quantities $k^{\mu\nu}$ being determined by the relations
\be \lb{kmn}
k^{\mu\nu} + \eta^{\mu \rho} \eta^{\nu\sigma}h_{\rho\sigma} + \eta^{\mu\rho} h_{\rho\sigma}k^{\nu\sigma} = 0.
\ee

According to equation (\ref{pmm}) the reception time transfer function may be written as 
\be \lb{Dr}
{\cal T}_{r}(\bx_A, t_B, \bx_B) = \frac{1}{c}R_{AB} + \frac{1}{c} \Delta_{r}(\bx_A, t_B, \bx_B),
\ee
where 
\be \lb{R}
R_{AB} = \vert \bx_{B} - \bx_{A}\vert
\ee
and $\Delta_{r}(\bx_A, t_B, \bx_B)$ is of the order of the gravitational perturbation $h_{\mu\nu}$. The function $\Delta_{r}/c$ defined by equation (\ref{Dr}) may be called the reception time delay function. It is well known that this quantity is $>0$ in Schwarzschild space-time, which explains its designation.

Replace now $\bx_{A}$ by a variable $\bx$ and consider $t_{B}, \bx_{B}$ as fixed parameters. Inserting ${\cal T}_{r}(\bx, t_B, \bx_B) = \vert \bx_{B} - \bx \vert /c + \Delta_{r}(\bx, t_B, \bx_B)/c$ into equation (\ref{eiko}) taken at $\bx$ instead of $\bx_{A}$, we get an equation which may be written in the form
\be \lb{del1}
2N^{i}\frac{\partial \Delta_{r}(\bx, t_{B}, \bx_{B})}{\partial x^{i}}  = - W(\bx, t_{B}, \bx_{B}),
\ee
where $N^{i}$ is defined as
\[
N^{i} = \frac{x_{B}^{i}-x^{i}}{\vert\bx_{B} - \bx\vert}
\]
and $W(\bx, t_{B}, \bx_{B})$ is given by
\bea
\fl W(\bx, t_{B}, \bx_{B}) &=& k^{00}(x_{-}) - 2 N^{i} k^{0i}(x_{-}) + N^{i} N^{j}k^{ij}(x_{-}) + 2\left[k^{0i}( x_{-}) - N^{j} k^{ij} (x_{-})\right]\nonumber\\
\fl&&\times\frac{\partial \Delta_{r}(\bx, t_{B}, \bx_{B})}{\partial x^{i}} + \left[\eta^{ij}+ k^{ij}(x_{-})\right]\left(\frac{\partial \Delta_{r}}{\partial x^{i}} \frac{\partial \Delta_{r}}{\partial x^{j}}\right)_{(\bx, t_B, \bx_{B})}, \lb{del2}
\eea
$x_{-}$ being the point-event defined by
\be \lb{xm}
x_{-} = (x_{B}^{0} - \vert \bx_{B} - \bx \vert - \Delta_{r}(\bx, t_B, \bx_{B}), \bx).
\ee

Since $\bx$ is a free variable, consider the case where $\bx$ is varying along the straight segment joining $\bx_A$ and $\bx_B$. Then we have
\be \lb{Nx}
N^{i} = N_{AB}^{i},
\ee
where $N_{AB}^{i}$ is by definition
\be \lb{Nab}
N_{AB}^{i} = \frac{x_B^i - x_A^i}{R_{AB}}, 
\ee  
and we can put
\be \lb{bxl}
\bx = \bz_{-}(\lambda),
\ee
where
\be \lb{bzl}
\bz_{-}(\lambda) = \bx_{B} - \lambda R_{AB} \bN_{AB}, \qquad 0\leq \lambda \leq 1.
\ee

A straightforward calculation shows that the total derivative of $\Delta_{r}(\bz_{-}(\lambda), t_B, \bx_{B})$ with respect to (w.r.t.) $\lambda$ is given by
\be \lb{der}
\frac{d}{d\lambda}\, \Delta_{r}(\bz_{-}(\lambda), t_B, \bx_{B}) = - R_{AB}N_{AB}^{i} \frac{\partial \Delta_{r}}{\partial x^{i}}(\bz_{-}(\lambda), t_B, \bx_B),
\ee
where $\partial  \Delta_{r}(\bz_{-}(\lambda), t_B, \bx_B)/\partial x^{i}$ denotes the partial derivative of $\Delta_{r}(\bx, t_B, \bx_B)$ w.r.t. $x^{i}$ taken at $\bx = \bz_{-}(\lambda)$.

Taking equation (\ref{Nx}) into account, and then comparing equation (\ref{der}) with equation (\ref{del1}), it may be seen that $\Delta_{r}(\bz_{-}(\lambda), t_B, \bx_{B})$ is governed by the differential equation 
\be \lb{del3}
\frac{d}{d\lambda} \,\Delta_{r}(\bz_{-}(\lambda), t_B, \bx_{B}) = \frac{1}{2} R_{AB} W(\bz_{-}(\lambda), t_{B}, \bx_{B})
\ee
with the boundary condition
\be \lb{bc1}
\Delta_{r}(\bz_{-}(0), t_B, \bx_{B}) = 0 
\ee
which follows from the obvious requirement $\Delta_{r}(\bx_B, t_B, \bx_B) = 0$ and from $\bx_B = \bz_{-}(0)$. As a consequence $\Delta_{r}(\bz_{-}(\lambda), t_B, \bx_{B})$ is such that
\be \lb{del4}
\Delta_{r}(\bz_{-}(\lambda), t_B, \bx_{B}) = \frac{1}{2} R_{AB}\int_{0}^{\lambda} W(\bz_{-}(\lambda'), t_{B}, \bx_{B})d \lambda'.
\ee

Noting that $\bz_{-}(1)= \bx_A$, we deduce a theorem as follows from Eqs. (\ref{del4}), (\ref{del2}) and (\ref{xm}). 

\medskip
{\bf Theorem 2.} {\em The reception time-delay function} $\Delta_{r}$ {\em defined by equation (\ref{Dr})} {\em satisfies the integro-differential equation}  
\bea
\fl\Delta_{r}(\bx_A, t_B, \bx_B) &=& \frac{1}{2} R_{AB}\int_{0}^{1}\bigg[ \left(k^{00} - 2 N_{AB}^{i} k^{0i} + N_{AB}^{i} N_{AB}^{j}k^{ij}\right)_{\widetilde{z}_{-}(\lambda)} \nonumber \\
\fl& &+ 2\left(k^{0i} - N_{AB}^{j} k^{ij}\right)_{\widetilde{z}_{-}(\lambda)} 
\frac{\partial \Delta_{r}}{\partial x^{i}}(\bz_{-}(\lambda), t_B, \bx_B)  \nonumber \\
\fl& &+ \left[\eta^{ij} + k^{ij}(\widetilde{z}_{-}(\lambda))\right]\left(\frac{\partial \Delta_{r}}{\partial x^{i}} \frac{\partial \Delta_{r}}{\partial x^{j}}\right)_{(\bz_{-}(\lambda), t_B, \bx_B)}\bigg] d\lambda , \lb{del5}
\eea
{\em where} $\widetilde{z}_{-}(\lambda)$ {\em is the point-event defined by}
\be \lb{tzm}
\widetilde{z}_{-}(\lambda) = (x_{B}^{0} - \lambda R_{AB} - \Delta_{r}(\bz_{-}(\lambda), t_B, \bx_B),\bz_{-}(\lambda)), 
\ee
$\bz_{-}(\lambda)$ {\em being given by equation (\ref{bzl})}.

\medskip
Of course, if we define the emission time delay function $\Delta_{e}(t_A, \bx_A, \bx_B)/c$ by the equation
\be \lb{De}
{\cal T}_{e}(t_A, \bx_A, \bx_B) = \frac{1}{c}R_{AB} + \frac{1}{c} \Delta_{e}(t_A, \bx_A, \bx_B).
\ee
a similar theorem may be stated for $\Delta_{e}$.

\medskip
{\bf Theorem 3.} {\em The emission time-delay function} $\Delta_{e}$ {\em defined by equation (\ref{De})} {\em satisfies the integro-differential equation} 
\bea
\fl\Delta_{e}(t_A, \bx_A,  \bx_B) &=& \frac{1}{2} R_{AB}\int_{0}^{1}\bigg[\left(k^{00} - 2 N_{AB}^{i} k^{0i} + N_{AB}^{i} N_{AB}^{j}k^{ij}\right)_{\widetilde{z}_{+}(\mu)} \nonumber \\
\fl& &- 2\left(k^{0i} - N_{AB}^{j} k^{ij}\right)_{\widetilde{z}_{+}(\mu)}\frac{\partial \Delta_{e}}{\partial x^{i}}(t_{A}, \bx_{A}, \bz_{+}(\mu))  \nonumber \\
\fl& &+ \left[\eta^{ij} + k^{ij}(\widetilde{z}_{+}(\mu))\right]\left(\frac{\partial \Delta_{e}}{\partial x^{i}} \frac{\partial \Delta_{e}}{\partial x^{j}}\right)_{(t_{A}, \bx_{A}, \bz_{+}(\mu))}\bigg] d\mu , \lb{dem1}
\eea
{\em where} $\widetilde{z}_{+}(\mu)$ {\em is the point-event defined by}
\be \lb{bzt2}
\widetilde{z}_{+}(\mu) = (x_{A}^{0} + \mu R_{AB} + \Delta_{e}(t_A, \bx_A, \bz_{+}(\mu)), \bz_{+}(\mu)) 
\ee    
{\em with}
\be \lb{bzm}
\bz_{+}(\mu) = \bx_{A} + \mu R_{AB} \bN_{AB},
\ee
{\em and} $\partial \Delta_{e}(t_A, \bx_A, \bz_{+}(\mu))/\partial x^i$ {\em denotes the partial derivative of} $\Delta_{e}(t_A, \bx_A, \bx)$ {\em w.r.t.} $x^i$ {\em taken at} $\bx = \bz_{+}(\mu)$.

\medskip
Integro-differential equations (\ref{del5}) and (\ref{dem1}) are very convenient to obtain the general post-Minkowskian expansion of each of the time transfer functions, as we shall see in the next section.

\section{General post-Minkowskian expansion of time delay functions}

Henceforth we suppose that each perturbation term $h_{\mu\nu}$ is represented at any point $x$ by a series in ascending powers of the Newtonian gravitational constant $G$
\be \lb{1n}
h_{\mu\nu}(x,G) =  \sum_{n=1}^{\infty}G^n \, g^{(n)}_{\mu\nu}(x).
\ee 
\noindent

The concomitant expansion of $k^{\mu\nu}$ is then given by 
\be \lb{1na} 
k^{\mu\nu}(x, G) = \sum_{n=1}^{\infty}G^n \, g^{\mu\nu}_{(n)}(x) \, ,
\ee
\noindent
where the set of quantities $g^{\mu\nu}_{(n)}$ can be recursively determined by using the relations
\be \lb{1n1}
g^{\mu\nu}_{(1)} = -\eta^{\mu\rho}  \eta^{\nu\sigma} g^{(1)}_{\rho\sigma} 
\ee
and
\be \lb{1nn}
g^{\mu\nu}_{(n)} = - \eta^{\mu\rho}  \eta^{\nu\sigma} g^{(n)}_{\rho\sigma} - 
\sum_{p=1}^{n-1}\eta^{\mu\rho} \, g^{(p)}_{\rho\sigma} \, g_{(n-p)}^{\nu\sigma}
\ee
\noindent
for $n\geq 2$.

As a consequence, the reception time delay function admits the expansion
\begin{equation} \label{PMt}
\Delta_{r}(\bx, t_B, \bx_B, G)=\sum_{n=1}^\infty G^n \Delta^{(n)}_r(\bx, t_B, \bx_B).
\end{equation}

It follows from equations (\ref{1na}) and (\ref{tzm}) that each term $k^{\mu\nu}$ involved in the right hand side of equation (\ref{del5}) may be written as
\be \lb{k-}
\fl k^{\mu\nu}(\widetilde{z}_{-}(\lambda), G) = \sum_{n=1}^{\infty}G^n g_{(n)}^{\mu \nu}(x_{B}^{0}- \lambda R_{AB}- \Delta_{r}(\bz_{-}(\lambda), t_B, \bx_B, G), \bz_{-}(\lambda)).
\ee

The general post-Minkowskian expansion of $k^{\mu\nu}(\widetilde{z}_{-}(\lambda), G)$ is obtained by substituting for $\Delta_{r}$ 
from equation (\ref{PMt}) into equation (\ref{k-}), and then performing Taylor's expansion about the point $z_{-}(\lambda)$ defined by
\be \lb{z0l}
z_{-}(\lambda) = (x_{B}^{0} - \lambda R_{AB}, \bz_{-}(\lambda)).
\ee
Let us put
\be \lb{Lk}
\fl\Phi_{r}^{(m,k)}(\bx, t_B, \bx_B) = \frac{(-1)^k}{k!}\sum_{l_1 + \cdots + l_k = m-k} \left\lbrack\prod_{j=1}^{k}
\Delta_{r}^{(l_j + 1)}(\bx, t_B, \bx_B)\right\rbrack ,
\ee
where $l_1, l_2, ..., l_k$ are either positive integers or zero ($m\geq 1$ and $1 \leq k \leq m$). A straightforward calculation yields
\be \lb{kex1}
k^{\mu\nu}(\widetilde{z}_{-}(\lambda),G) = \sum_{n=1}^{\infty} G^{n} \widehat{g}_{- (n)}^{\mu \nu}(z_{-}(\lambda), t_B, \bx_B),
\ee
where the quantities $\widehat{g}_{- (n)}^{\mu \nu}(z_{-}(\lambda), t_B, \bx_B)$ are given by
\be \lb{hg1}
\widehat{g}_{- (1)}^{\mu \nu}(z_{-}(\lambda), t_B, \bx_B) = g_{(1)}^{\mu \nu}(z_{-}(\lambda))
\ee
and 
\bea 
\fl\widehat{g}_{- (n)}^{\mu \nu}(z_{-}(\lambda), t_B, \bx_B) &=& g_{(n)}^{\mu \nu}(z_{-}(\lambda)) \nonumber\\
\fl\lb{hgn}&&+ \sum_{m=1}^{n-1} \sum_{k=1}^{m}
\Phi_{r}^{(m,k)}(\bz_{-}(\lambda), t_B, \bx_B) \left[\frac{\partial^{k} g_{(n-m)}^{\mu \nu}}{(\partial x^{0})^{k}}\right]_{z_{-}(\lambda)}
\eea
for $n\geq 2$. Substituting for $k^{\mu\nu}(\widetilde{z}_{-}(\lambda),G)$ from equation (\ref{kex1}) into equation (\ref{del5}), we get the theorem which follows.

\medskip
{\bf Theorem 4.} {\em When the perturbation term} $h_{\mu\nu}$ {\em in the metric is represented by expansion (\ref{1n}), the reception time-delay function} $\Delta_{r}$ {\em is given by equation (\ref{PMt}), namely}
\[
\Delta_{r}(\bx, t_B, \bx_B, G)=\sum_{n=1}^\infty G^n \Delta^{(n)}_r(\bx, t_B, \bx_B),
\]
{\em where}  
\bea
\fl\Delta_{r}^{(1)}(\bx_A, t_B, \bx_B) &=& \frac{1}{2}R_{AB}\int_{0}^{1}\left[g_{(1)}^{00} - 2 N_{AB}^{i} g_{(1)}^{0i} + N_{AB}^{i} N_{AB}^{j}g_{(1)}^{ij}\right]_{z_{-}(\lambda)} d\lambda \, ,\lb{Tr1} \\ 
\fl\Delta_{r}^{(2)}(\bx_A, t_B, \bx_B) &=& \frac{1}{2}R_{AB}\int_{0}^{1}\bigg\{ \left[\widehat{g}_{-(2)}^{00} - 2 N_{AB}^{i} \widehat{g}_{-(2)}^{0i} + N_{AB}^{i} N_{AB}^{j}\widehat{g}_{-(2)}^{ij}\right]_{(z_{-}(\lambda), t_B, \bx_B)} \nonumber \\
\fl& &+ 2\left[g_{(1)}^{0i} - N_{AB}^{j} g_{(1)}^{ij}\right]_{z_{-}(\lambda)} \frac{\partial \Delta_{r}^{(1)}}{\partial x^{i}}(\bz_{-}(\lambda), t_B, \bx_B) \nonumber\\
\fl&&+ \eta^{ij}\left[\frac{\partial \Delta_{r}^{(1)}}{\partial x^{i}} \frac{\partial \Delta_{r}^{(1)}}{\partial x^{j}}\right]_{(\bz_{-}(\lambda), t_B, \bx_B)}\bigg\} d\lambda  \lb{Tr2}
\eea
{\em and} 
\bea
\fl\Delta_{r}^{(n)}(\bx_A, t_B, \bx_B) &=&\frac{1}{2}R_{AB}\int_{0}^{1}\bigg\{ \left[\widehat{g}_{-(n)}^{00} - 2 N_{AB}^{i} \widehat{g}_{-(n)}^{0i} + N_{AB}^{i} N_{AB}^{j}\widehat{g}_{-(n)}^{ij}\right]_{(z_{-}(\lambda), t_B, \bx_B)} \nonumber \\
\fl& &+ 2\sum_{p=1}^{n-1}\left[\widehat{g}_{-(p)}^{0i} - N_{AB}^{j} \widehat{g}_{-(p)}^{ij}\right]_{(z_{-}(\lambda), t_B, \bx_B)} \frac{\partial \Delta_{r}^{(n-p)}}{\partial x^{i}}(\bz_{-}(\lambda), t_B, \bx_B)\nonumber \\
\fl& & + \sum_{p=1}^{n-1}\eta^{ij}\left[\frac{\partial \Delta_{r}^{(p)}}{\partial x^{i}} \frac{\partial \Delta_{r}^{(n-p)}}{\partial x^{j}}\right]_{(\bz_{-}(\lambda), t_B, \bx_B)}\nonumber \\
\fl& &+\sum_{p=1}^{n-2}\widehat{g}_{-(p)}^{ij}(z_{-}(\lambda), t_B, \bx_B)\nonumber\\
\fl&&\qquad\quad\times \sum_{q=1}^{n-p-1} \left[\frac{\partial \Delta_{r}^{(q)}}{\partial x^{i}} \frac{\partial \Delta_{r}^{(n-p-q)}}{\partial x^{j}}\right]_{(\bz_{-}(\lambda), t_B, \bx_B)}\bigg\} d\lambda  \lb{Trn}
\eea
{\em for }$n\geq 3${\em , the quantities} $\widehat{g}_{- (n)}^{\mu \nu}$ {\em being defined by equations (\ref{hg1}) and (\ref{hgn})}. 

\medskip
It may be noted that the integral expressions occurring in Eqs. (\ref{Tr1})-(\ref{Trn}) are line integrals taken along the zeroth-order null geodesic of parametric equation $x = z_{-}(\lambda)$, where $z_{-}(\lambda)$ is defined by equation (\ref{z0l}). 

\medskip
\par A similar reasoning works for the emission time transfer function, which admits the expansion 
\be \lb{td2}
\Delta_{e}(t_A, \bx_A, \bx, G) = \sum_{n=1}^{\infty} G^n \Delta_{e}^{(n)}(t_A, \bx_A, \bx).
\ee

We define the functions $\Phi_{e}^{(m,k)}(t_A, \bx_A, \bx)$ as 
\be \lb{Lk2}
\Phi_{e}^{(m,k)}(t_A, \bx_A, \bx) = \frac{1}{k!}\sum_{l_1 + \cdots + l_k = m-k} \left\lbrack\prod_{j=1}^{k}
\Delta_{e}^{(l_j + 1)}(t_A, \bx_A, \bx)\right\rbrack ,
\ee
where $l_1, l_2, ..., l_k$ are either positive integers or zero. Then, defining $z_{+}(\mu)$ as the point-event
\be \lb{bz3}
z_{+}(\mu) = (x_{A}^{0} + \mu R_{AB}, \bz_{+}(\mu)), 
\ee
where $\bz_{+}(\mu)$ is given by equation (\ref{bzm}), we put
\be \lb{gh1}
\widehat{g}_{+ (1)}^{\mu \nu}(t_A, \bx_A, z_{+}(\mu)) = g_{(1)}^{\mu \nu}(z_{+}(\mu))
\ee
and 
\bea 
\fl\widehat{g}_{+ (n)}^{\mu \nu}( t_A, \bx_A, z_{+}(\mu)) &=& g_{(n)}^{\mu \nu}(z_{+}(\mu)) \nonumber\\
&&+ \sum_{m=1}^{n-1} \sum_{k=1}^{m}
\lb{ghn}\Phi_{e}^{(m,k)}(t_A, \bx_A, \bz_{+}(\mu)) \left[\frac{\partial^{k} g_{(n-m)}^{\mu \nu}}{(\partial x^{0})^{k}}\right]_{z_{+}(\mu)}
\eea
for $n\geq 2$. Thus we can formulate a theorem as follows.

\medskip
{\bf Theorem 5.} {\em On the asumption of Theorem 4, the emission time transfer function is given by equation (\ref{td2}), where}
\bea
\fl\Delta_{e}^{(1)}(t_A, \bx_A, \bx_B) &=& \frac{1}{2}R_{AB}\int_{0}^{1}\left[g_{(1)}^{00} - 2 N_{AB}^{i} g_{(1)}^{0i} + N_{AB}^{i} N_{AB}^{j}g_{(1)}^{ij}\right]_{z_{+}(\mu)} d\mu \, ,\lb{Te1} \\ 
\fl\Delta_{e}^{(2)}(t_A, \bx_A, \bx_B) &=& \frac{1}{2}R_{AB}\int_{0}^{1}\bigg\{ \left[\widehat{g}_{+(2)}^{00} - 2 N_{AB}^{i} \widehat{g}_{+(2)}^{0i} + N_{AB}^{i} N_{AB}^{j}\widehat{g}_{+(2)}^{ij}\right]_{(t_A, \bx_A, z_{+}(\mu))} \nonumber \\
\fl& &- 2\left[g_{(1)}^{0i} - N_{AB}^{j} g_{(1)}^{ij}\right]_{z_{+}(\mu)} \frac{\partial \Delta_{e}^{(1)}}{\partial x^{i}}(t_A, \bx_A, \bz_{+}(\mu)) \nonumber\\
\fl &&+ \eta^{ij}\left[\frac{\partial \Delta_{e}^{(1)}}{\partial x^{i}} \frac{\partial \Delta_{e}^{(1)}}{\partial x^{j}}\right]_{(t_A, \bx_A, \bz_{+}(\mu))}\bigg\} d\mu  \lb{Te2}
\eea
{\em and} 
\bea
\fl&&\Delta_{e}^{(n)}(t_A, \bx_A, \bx_B) =\frac{1}{2}R_{AB}\int_{0}^{1}\bigg\{ \left[\widehat{g}_{+(n)}^{00} - 2 N_{AB}^{i} \widehat{g}_{+(n)}^{0i} + N_{AB}^{i} N_{AB}^{j}\widehat{g}_{+(n)}^{ij}\right]_{(t_A, \bx_A, z_{+}(\mu))} \nonumber \\
\fl& &\qquad\quad- 2\sum_{p=1}^{n-1}\left[\widehat{g}_{+(p)}^{0i} - N_{AB}^{j} \widehat{g}_{+(p)}^{ij}\right]_{(t_A, \bx_A, z_{+}(\mu))} \frac{\partial \Delta_{e}^{(n-p)}}{\partial x^{i}}(t_A, \bx_A, \bz_{+}(\mu))\nonumber \\
\fl& &\qquad\quad + \sum_{p=1}^{n-1}\eta^{ij}\left[\frac{\partial \Delta_{e}^{(p)}}{\partial x^{i}} \frac{\partial \Delta_{e}^{(n-p)}}{\partial x^{j}}\right]_{(t_A, \bx_A, \bz_{+}(\mu))}\nonumber \\
\fl& &\qquad\quad+\sum_{p=1}^{n-2}\widehat{g}_{+(p)}^{ij}(t_A, \bx_A, z_{+}(\mu)) \sum_{q=1}^{n-p-1} \left[\frac{\partial \Delta_{e}^{(q)}}{\partial x^{i}} \frac{\partial \Delta_{e}^{(n-p-q)}}{\partial x^{j}}\right]_{(t_A, \bx_A, \bz_{+}(\mu))}\bigg\} d\mu \lb{Ten}
\eea
{\em for} $n\geq 3${\em , the quantities} $\widehat{g}_{+ (n)}^{\mu \nu}$ {\em being defined by equations (\ref{gh1}) and (\ref{ghn})}. 

\medskip
The integrals are now taken along the zeroth-order null geodesic defined by the parametric equation $x= z_{+}(\mu)$.

\medskip
As far as we know, the above results are new for $n\geq 2$. On the other hand, the expressions obtained here for $\Delta_{r}^{(1)}$ and $\Delta_{e}^{(1)}$ are equivalent to the well-known expressions derived by other methods. Taking equation (\ref{1n1}) into account, it may be seen that the formulae (\ref{Tr1}) and (\ref{Te1}) coincide with the expressions obtained, e.g., in Ref. \cite{2004CQGra..21.4463L} for $c{\cal T}_{r}^{(1)}$ and $c{\cal T}_{e}^{(1)}$, respectively.    

\section{Case of a stationary space-time}

In the case of a stationary space-time, one can choose coordinates $(x^{\alpha})$ such that the components of the metric do not depend on $x^{0}$. Then the two time transfer functions reduce to a single one, so that we can write:
\be \lb{T}
\mathcal{T}_r(\bx_A, t_B, \bx_B) \equiv \mathcal{T}_e(t_A,\bx_A,\bx_B) \equiv \mathcal{T}(\bx_A,\bx_B).
\ee

As a consequence, the functions $\Delta_{r}$ and $\Delta_{e}$ are identical and depend only on $\bx_{A}$ and $\bx_{B}$, which implies that
\be \lb{Dre}
\Delta_{r}^{(n)}(\bx_{A}, t_{B}, \bx_{B}) \equiv \Delta_{e}^{(n)}(t_{A}, \bx_{A}, \bx_{B}) \equiv \Delta^{(n)}(\bx_{A}, \bx_{B})
\ee 
for any $n \geq 1$. 

Moreover, the stationary character of the metric implies that 
\be \lb{g-}
\widehat{g}_{-(n)}^{\mu \nu}(z_{-}(\lambda), t_B, \bx_B) \equiv g_{(n)}^{\mu \nu}(\bz_{-}(\lambda))   
\ee
and 
\be \lb{g+}
\widehat{g}_{+(n)}^{\mu \nu}(t_A, \bx_A, z_{+}(\mu)) \equiv g_{(n)}^{\mu \nu}(\bz_{+}(\mu)).
\ee

So equations (\ref{Tr1}) -(\ref{Trn}) and (\ref{Te1})-(\ref{Ten}) simplify. For example, equations (\ref{Te1}) and (\ref{Te2}) may be written as 
\be \lb{D1}
\fl\Delta^{(1)}(\bx_A, \bx_B) = \frac{1}{2}R_{AB}\int_{0}^{1}\left[g_{(1)}^{00} - 2 N_{AB}^{i} g_{(1)}^{0i} + N_{AB}^{i} N_{AB}^{j}g_{(1)}^{ij}\right]_{\bz_{+}(\mu)} d\mu  
\ee
and
\bea 
\fl&&\Delta^{(2)}(\bx_A, \bx_B) = \frac{1}{2}R_{AB}\int_{0}^{1}\bigg\{ \left[g_{(2)}^{00} - 2 N_{AB}^{i} g_{(2)}^{0i} + N_{AB}^{i} N_{AB}^{j} g_{(2)}^{ij}\right]_{\bz_{+}(\mu)} \nonumber \\
\fl& &- 2\left[g_{(1)}^{0i} - N_{AB}^{j} g_{(1)}^{ij}\right]_{\bz_{+}(\mu)} \frac{\partial \Delta^{(1)}}{\partial x^{i}}(\bx_A, \bz_{+}(\mu)) + \eta^{ij}\left[\frac{\partial \Delta^{(1)}}{\partial x^{i}} \frac{\partial \Delta^{(1)}}{\partial x^{j}}\right]_{(\bx_A, \bz_{+}(\mu))}\bigg\} d\mu, \nonumber\\
\fl &&\lb{D2}
\eea
respectively. 

A straightforward calculation shows that making the change of variable
\[
\lambda = - \mu
\]
transforms the right hand side of equations (\ref{D1}) and (\ref{D2}) into the expressions deduced from equations (\ref{Tr1}) and (\ref{Trn}), respectively when equations (\ref{Dre}) and (\ref{g-}) are taken into account. It is worthy of note that the integrals reduce in this case to line integrals taken along the segment joining $\bx_{A}$ and $\bx_{B}$.

\section{Application to a static, spherically symmetric space-time}

To illustrate the previous results, let us consider the gravitational field outside a static, spherically symmetric body of mass $M$. Choosing spatial quasi Cartesian isotropic coordinates and putting $r= \vert \bx \vert$, we suppose that the metric has the form
\bea
ds^2 &=& \left(1 - \frac{2G M}{c^2 r} + 2\beta \frac{G^2 M^2}{c^4 r^2} + ...\right)\left(dx^{0}\right)^2 \nonumber\\
\lb{sss}&&\qquad\quad- \left(1 + 2 \gamma \frac{G M}{c^2 r} + \frac{3}{2}\delta \frac{G^2 M^2}{c^4 r^2} + ...\right)\delta_{ij}dx^{i} dx^{j},
\eea
where $\beta$ and $\gamma$ are the usual post-Newtonian parameters, $\delta$ is a post-post-Newtonian parameter and $+...$ means that terms of order $G^3$ are neglected ($\beta=\gamma =\delta =1$ in general relativity). Furthermore, we suppose that points $x_A$ and $x_B$ are such that the geodesic path connecting them is entirely outside the body. We use the notations
\[ 
r_A = \vert \bx_A \vert \, , \quad  r_B  = \vert \bx_B \vert.
\]

The contravariant components of the metric are given by
\begin{eqnarray}
\fl g_{(1)}^{00}&=& \frac{2M}{c^2r}, \; \qquad \qquad \quad \; \; g_{(1)}^{0i} = 0, \; \qquad g_{(1)}^{ij}= 2\gamma\frac{M}{c^2r}\delta^{ij} \, , \lb{ssp1} \\
\fl g_{(2)}^{00}&=& 2(2 -\beta)\frac{M^2}{c^4r^2}, \qquad \; \, g_{(2)}^{0i} = 0, \qquad \; g_{(2)}^{ij}=(\frac{3}{2}\delta -4\gamma^2)\frac{M^2}{c^4r^2}\delta^{ij}. \lb{ssp2}
\end{eqnarray}

Taking equations (\ref{ssp1}) into account, we immediately deduce from equation (\ref{D1}) that $\Delta^{(1)}$ is given by the well-known formula
\be \lb{sha}
\fl\Delta^{(1)}(\bx_A, \bx_B) = \frac{(\gamma+1)M}{c^2}R_{AB}\int_{0}^{1}\frac{d\mu}{\vert \bz_{+}(\mu)\vert} = \frac{(\gamma+1)M}{c^2}\ln \left( \frac{r_A+r_B+R_{AB}}{r_A+r_B-R_{AB}}\right),
\ee
which is equivalent to the expression of the time delay found by Shapiro \cite{shapiro}.

Substituting now equations (\ref{ssp1}) and (\ref{ssp2}) into equation (\ref{D2}), we get 
\bea 
\fl&&\Delta^{(2)}(\bx_A, \bx_B) = \frac{1}{4}R_{AB}\int_{0}^{1}\bigg\lbrace (8 - 4\beta -8\gamma^{2} + 3\delta)\frac{M^{2}}{c^{4} \vert \bz_{+}(\mu)\vert^{2}} \nonumber \\
\fl&&\qquad + 8\gamma \frac{M}{c^{2}\vert\bz_{+}(\mu)\vert} N_{AB}^{i} \frac{\partial \Delta^{(1)}}{\partial x^{i}}(\bx_A, \bz_{+}(\mu)) 
- 2\sum_{i=1}^{3}\left(\frac{\partial \Delta^{(1)}}{\partial x^{i}}(\bx_A, \bz_{+}(\mu))\right)^{2}\bigg\rbrace d\mu. \nonumber \\
\fl&& \lb{sD2}
\eea

We deduce from equation (\ref{sha}) that 
\be \lb{ND}
N_{AB}^{i} \frac{\partial \Delta^{(1)}}{\partial x^{i}}(\bx_A, \bz_{+}(\mu)) = (\gamma + 1)\frac{M}{c^{2} \vert\bz_{+}(\mu)\vert}
\ee
and
\bea 
\fl&&\sum_{i=1}^{3}\left(\frac{\partial \Delta^{(1)}}{\partial x^{i}}(\bx_A, \bz_{+}(\mu))\right)^{2} = 2 (\gamma + 1)^{2} \frac{M^{2}}{c^{4}}
\frac{d}{d\mu}\left[\frac{\mu}{r_{A} \vert\bz_{+}(\mu)\vert + \mu \bx_A . (\bx_B - \bx_A) + r_{A}^{2}} \right]. \nonumber \\
\fl&& \lb{DD}
\eea

Substituting equations (\ref{ND}) and (\ref{DD}) into equation (\ref{sD2}), and then noting that
\be \lb{sE}
\int_{0}^{1}\frac{d\mu}{\vert \bz_{+}(\mu)\vert^{2}} = \frac{\arccos (\bn_{A}. \bn_{B})}{r_{A} r_{B} \sqrt{1 - (\bn_{A}. \bn_{B})^{2}}},
\ee
where 
\be \lb{nAB}
\bn_{A} = \frac{\bx_{A}}{r_{A}}, \qquad \bn_{B} = \frac{\bx_{B}}{r_{B}},
\ee
we obtain by a straightforward calculation
\bea 
\fl&&\Delta^{(2)}(\bx_A, \bx_B) = \frac{M^{2}}{c^{4}}\frac{R_{AB}}{r_{A} r_{B}}\left[
\frac{(8-4\beta+8\gamma+3\delta)\arccos (\bn_A . \bn_B)}
{4\sqrt{1-(\bn_{A}. \bn_{B})^2}}
-\frac{(1+\gamma)^2}{1 + (\bn_{A} . \bn_{B})}\right]. \nonumber \\
\fl&& \lb{D2b}
\eea

Finally, using the formulae (\ref{sha}) and (\ref{D2b}), we get an expression as follows for the time transfer function
\begin{eqnarray}
\fl\mathcal{T}(\bx_A,\bx_B) &=& \frac{R_{AB}}{c} +\frac{(\gamma+1)GM}{c^{3}}
\ln\left(\frac{r_A+r_B+R_{AB}}{r_A+r_B-R_{AB}}\right)\nonumber \\
\fl& & +\frac{G^{2}M^{2}}{c^{5}}\frac{R_{AB}}{r_{A} r_{B}}\left[
\frac{(8-4\beta+8\gamma+3\delta)\arccos (\bn_A . \bn_B)}
{4\sqrt{1-(\bn_{A}. \bn_{B})^2}}
-\frac{(1+\gamma)^2}{1 + (\bn_{A} . \bn_{B})}\right]\nonumber\\
\fl&& + O(G^3).  \lb{7n}
\end{eqnarray}

We recover a result previously derived by different approaches \cite{richter2,2004CQGra..21.4463L} (see also Refs. \cite{john} and \cite{brumberg} in the case where $\beta = \gamma = \delta = 1$).

\section{Concluding remarks}
\par Equations (\ref{Tr1})-(\ref{Trn}) and (\ref{Te1})-(\ref{Ten}) are the main results of this paper. These equations show that 
the time transfer functions can be obtained within the $n$th post-Minkowskian approximation by a recursive procedure which spares 
the trouble of solving the geodesic differential equations and avoids determining Synge's world function. It is remarkable that any $n$th-order perturbation term is an integral taken along a zeroth-order null straight line. The derivation of the time transfer function performed here for a static, spherically symmetric space-time is significantly more simple than the calculation carried out in Ref. \cite{2004CQGra..21.4463L}.  

As a final remark, it may be noted that since the time transfer functions are sufficient to determine the deflection 
and the frequency shift of a light signal, the systematic method developed here will be very convenient to tackle the 
relativistic problems raised by highly accurate astrometry and time/frequency metrology.


\section*{References}
\begin{thebibliography}{99}
\bibitem{turyshev} 
Turyshev S G, Shao M and Nordtvedt K L 2008 Science, Technology, and Mission Design for the Laser Astrometric Test of Relativity {\em Lasers, Clocks and Drag-Free Control: Exploration of Relativistic Gravity in Space} ({\em Springer Series in Astrophysics and Space Science Library} vol {\bf 349}) 
ed H Dittus, C L\"{a}mmerzahl and S G Turyshev (Berlin: Springer) p 473 ({\em Preprint} gr-qc/0601035)
\bibitem{Ni}
Ni W T 2007 {\em Preprint} gr-qc/0712.2492
\bibitem{will}
Will C M 1993 {\em Theory and Experiment in Gravitational Physics} (Cambridge: Cambridge University Press) 2nd edition
\bibitem{blanchet}
Blanchet L, Salomon C, Teyssandier P and Wolf P 2001 {\em Astron. Astrophys.} {\bf 370} 320 
\bibitem{Kopeikin:1997}
Kopeikin S M 1997 {\em J. Math. Phys.} (N.Y.) {\bf 38} 2587 
\bibitem{kopeikin2} 
Kopeikin S M and Sch\"{a}fer G 1999 {\em Phys. Rev.} D {\bf 60} 124002
\bibitem{kopeikin3} 
Kopeikin S M and Mashhoon B 2002 {\em Phys. Rev.} D {\bf 65} 064025
\bibitem{2003PhLA..308..101C} 
Ciufolini I, Kopeikin S, Mashhoon B and Ricci F 2003 {\em Phys. Lett.} A {\bf 308} 101 
\bibitem{richter1} 
Richter G W and Matzner R A 1982 {\em Phys. Rev.} D {\bf 26} 2549 
\bibitem{richter2} 
Richter G W and Matzner R A 1983 {\em Phys. Rev.} D {\bf 28} 3007 
\bibitem{john}
John R W 1975 {\em Exp. Tech. Physik} {\bf 23} 127 
\bibitem{brumberg}
Brumberg V A 1987 {\em Kinematics Phys. Celest. Bodies} {\bf 3} 6 
\bibitem{2002PhRvD..66b4045L} 
Linet B and Teyssandier P 2002 {\em Phys. Rev.} D {\bf 66} 024045 
\bibitem{2004CQGra..21.4463L}  
Le Poncin-Lafitte C, Linet B and Teyssandier P 2004 {\em Class. Quantum Grav.} {\bf 21} 4463 
\bibitem{universal}
Teyssandier P, Le Poncin-Lafitte C and Linet B 2008 A Universal Tool for Determining the Time Delay and the Frequency Shift 
of Light: Synge's World Function {\em Lasers, Clocks and Drag-Free Control: Exploration of Relativistic Gravity in Space} ({\em Springer Series in Astrophysics and Space Science Library} vol {\bf 349}) ed H Dittus, C L\"{a}mmerzahl and S G Turyshev (Berlin: Springer) p 153 ({\em Preprint} gr-qc/0711.0034)
\bibitem{synge1}  
Synge J L 1964 {\em Relativity: The General Theory} (Amsterdam: North-Holland)
\bibitem{bel}
Bel Ll and Mart\'{i}n J 1994 {\em Gen. Rel. Grav.} {\bf 26} 567 
\bibitem{shapiro}
Shapiro I I 1964 {\em Phys. Rev. Lett.} {\bf 13} 789 
\end{thebibliography}
\end{document}